\documentclass[aps,prl,twocolumn,superscriptaddress,showpacs]{revtex4-1}
\usepackage{graphicx} % for setting figures
\usepackage{wasysym} % for circulating state symbols

\begin{document}

\title{Heralded state preparation in a superconducting qubit}

\author{J. E. Johnson}
\affiliation{Department of Physics, University of California, Berkeley, California 94720, USA}
\author{C. Macklin}
\affiliation{Quantum Nanoelectronics Laboratory, Department of Physics, University of California, Berkeley, California 94720, USA}
\author{D. H. Slichter}
\affiliation{Quantum Nanoelectronics Laboratory, Department of Physics, University of California, Berkeley, California 94720, USA}
\author{R. Vijay}
\affiliation{Quantum Nanoelectronics Laboratory, Department of Physics, University of California, Berkeley, California 94720, USA}
\author{E. B. Weingarten}
\affiliation{Department of Physics, University of California, Berkeley, California 94720, USA}
\author{John Clarke}
\affiliation{Department of Physics, University of California, Berkeley, California 94720, USA}
\author{I. Siddiqi}
\affiliation{Quantum Nanoelectronics Laboratory, Department of Physics, University of California, Berkeley, California 94720, USA}

%\thanks{}

\date{\today}

\begin{abstract}
We demonstrate high-fidelity, quantum nondemolition, single-shot readout of a superconducting flux qubit in which the pointer state distributions can be resolved to below one part in 1000. In the weak excitation regime, continuous measurement permits the use of heralding to ensure initialization to a fiducial state, such as the ground state. This procedure boosts readout fidelity to 93.9\% by suppressing errors due to spurious thermal population. Furthermore, heralding potentially enables a simple, fast qubit reset protocol without changing the system parameters to induce Purcell relaxation.
\end{abstract}


\maketitle

Recent progress in superconducting qubits \cite{Clarke:2008,Neeley:2010uq,DiCarlo:2010kx,Irfan_review} has resulted in coherence times exceeding 10 $\mu$s \cite{Bylander:2011,3D_Transmon} as well as the demonstration of multiqubit algorithms \cite{DiCarlo:2009, Martinis_vonNeumann}, further validating superconducting circuits as a viable platform for quantum information processing.  The simultaneous realization of a fast, high-fidelity, quantum nondemolition (QND) qubit readout \cite{Lupascu:2007}, however, has thus far been difficult, with many schemes exhibiting either sub-unity visibility \cite{ReedOut}, long measurement times \cite{PhysRevB.73.054510,Mallet:2009oq}, or demolition of the quantum state \cite{Martinis_phasequbit}. Low power dispersive readouts based on circuit quantum electrodynamics (cQED) \cite{PhysRevA.69.062320,Wallraff_nature} have demonstrated QND operation \cite{QND_singlephoton} with near unity visibility, thus faithfully mapping the qubit state to distinct frequency shifts of a microwave resonator \cite{unit_visibility}. Traditionally, insufficient measurement sensitivity has limited the fidelity with which these pointer states can be resolved in this architecture. Recent advances in low-noise superconducting amplifiers \cite{Muck:1998,10.1063/1.3486156,J_amp_2011,Spietz:2009ly,Castellanos-Beltran:2007ve,McDermott_slug} have provided a sufficiently high signal-to-noise ratio (SNR) \cite{Vijay:2011bh,Johnson_fluxqubit} to resolve the state of the measurement cavity in a time much shorter than the qubit relaxation time $T_1$, and thus achieve single-shot readout.

In this Letter, we use a high speed readout based on a Josephson parametric amplifier (paramp) \cite{Hatridge:2011qf} to insert an additional measurement pulse before a qubit manipulation and measurement sequence to verify that the quantum system is initialized in the ground state. Such heralding techniques are currently employed in other quantum information architectures such as trapped ions \cite{heralding_optics}, photonic systems \cite{heralding_optics_II}, and quantum dots \cite{quantum_dot_heralded}. With this technique, we effectively eliminate state preparation errors due to the spurious excited state population observed in superconducting qubits \cite{Palacios_thermal,Steffen_CSFQ}. Furthermore, this method permits a rapid, deterministic reset of the qubit state\textemdash a particularly important function in long-lived qubits where simply waiting for a time much longer than $T_1$ is impractical.

\begin{figure}
\includegraphics[width=8.6cm]{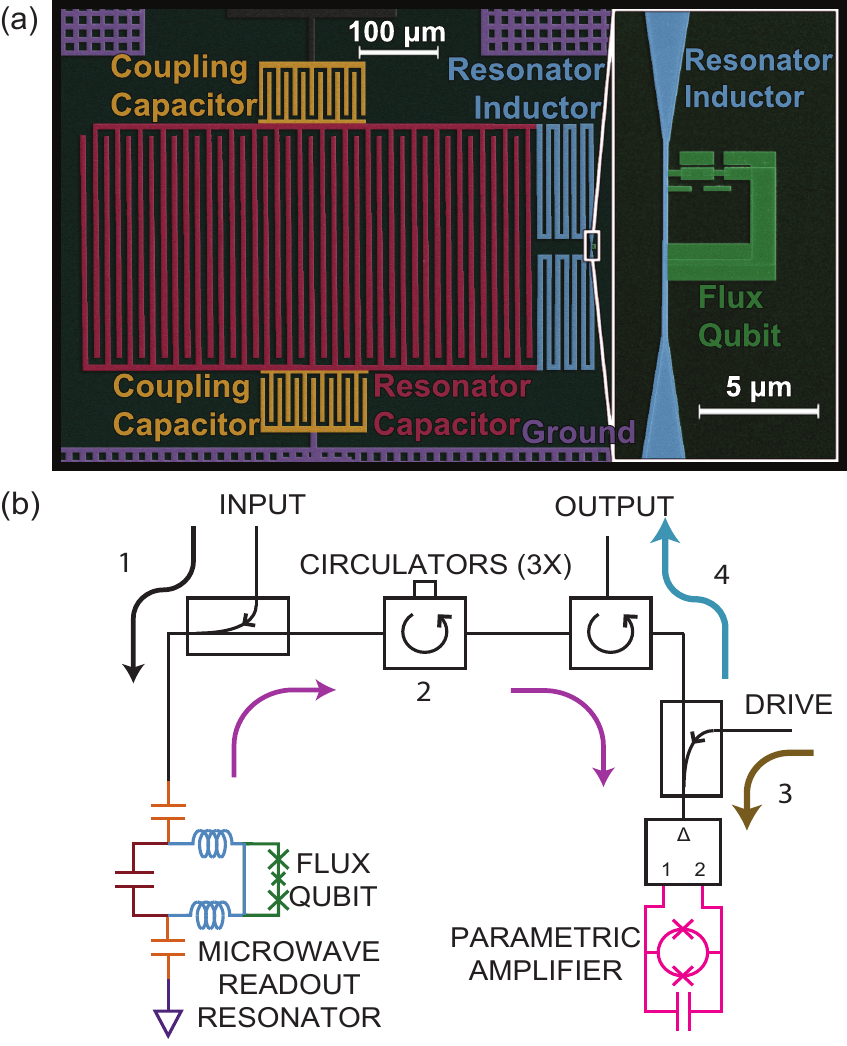}
\caption{(Color online) Experimental setup. (a) False-color SEM image of the qubit and readout resonator with a magnified inset of the qubit. (b) Path of the readout signal at 30 mK.}
\label{fig:falsecolor}
\end{figure}

The flux qubit and quasi-lumped-element measurement resonator [Fig. \ref{fig:falsecolor}(a)] are fabricated from aluminum on a silicon substrate using double-angle shadow evaporation. The resonator is formed by the parallel combination of an interdigitated finger capacitor ($C_r = 600$~fF) and meander line inductor ($L_r = 1.3$~nH), providing a measured resonant frequency $\omega_r/ 2 \pi = 5.780$~GHz. The resonator is coupled to the 50-$\Omega$ feed line by planar coupling capacitors, which set the resonator linewidth $\kappa/ 2 \pi = 9$~MHz. A perforated metal plane surrounds the resonator and forms the ground reference for an on-chip, coplanar waveguide input for both the qubit manipulation and readout pulses.

The three-junction flux qubit (loop dimensions $3.8 \times 4.0~ \mu$m$^2$) is inductively coupled to the resonator through a shared, 2.6-$\mu$m length of a 150-nm-wide constriction in the meander line inductor.  The qubit is flux biased with a small superconducting coil mounted on the outside of the copper box containing the qubit chip. The cryopackage is anchored to the 30 mK base temperature stage of a liquid-cryogen-free dilution refrigerator. A schematic of the cryogenic portion of the microwave measurement circuit is shown in Fig. \ref{fig:falsecolor}(b). Qubit control and readout pulses are injected through the weakly coupled port of a directional coupler \footnote{Manufactured by Krytar, part no. 104020020}.  The readout pulse is reflected from the resonator, acquiring a qubit-state-dependent phase shift. The reflected pulse passes through a series of circulators \footnote{Manufactured by Pamtech, part no. CTH1368K18} and a second directional coupler to the paramp, where it is amplified and reflected to the output port. The readout signal is further amplified by cryogenic and room-temperature amplifiers before it is down-converted to zero frequency and digitized in 10-ns increments.

The paramp is flux biased to match the qubit readout frequency and operated in phase-sensitive mode, wherein a strong pump tone is applied at the same frequency as the readout pulse. In this operating mode, the phase of the reflected pump depends very sensitively on its power. If a small input signal is added in phase with the pump signal, the resulting small change in net power leads to a large phase shift. This form of amplification theoretically adds no additional noise, allowing the system noise temperature to remain close to the standard quantum limit $T_Q = \hbar \omega /2 k_B = 139$~mK \cite{Hatridge:2011qf}. Our paramp provides 20-dB gain with an instantaneous bandwidth of 7 MHz.

We use spectroscopy to determine the flux qubit parameters $\Delta/h = 6.15$~GHz (energy splitting at the degeneracy point) and $I_q = 204$~ nA (circulating supercurrent in the qubit loop) \cite{vanderWal2000}. The coupling strength between the qubit and resonator is $g/2\pi = 105$~ MHz, from which we extract the qubit-resonator mutual inductance $M_t = 6.3$~pH. In this geometry, the effective coupling strength is given by $g'=g\sin(\theta)$, where $\sin(\theta) = \Delta/\hbar\omega_{01}$; $\omega_{01}/2\pi$ is the qubit transition frequency \cite{Abdumalikov}. We flux bias the qubit away from the degeneracy point at $\omega_{01}/2\pi = 7.80$~GHz because the small qubit-resonator detuning otherwise induces a strong nonlinearity in the resonator. The operating bias corresponds to a detuning $\delta/2 \pi = 2.02$~GHz and a dispersive shift $2 \chi/2\pi = 7.4$~MHz, the latter producing a 150$^{\circ}$ phase shift between the two qubit pointer states. The relaxation time is $T_1 = 1.8 \ \mu$s, while the Rabi and Ramsey ($T_2^*$) decay times are 500 ns and 55 ns, respectively.

\begin{figure}
\includegraphics[width=8.6cm]{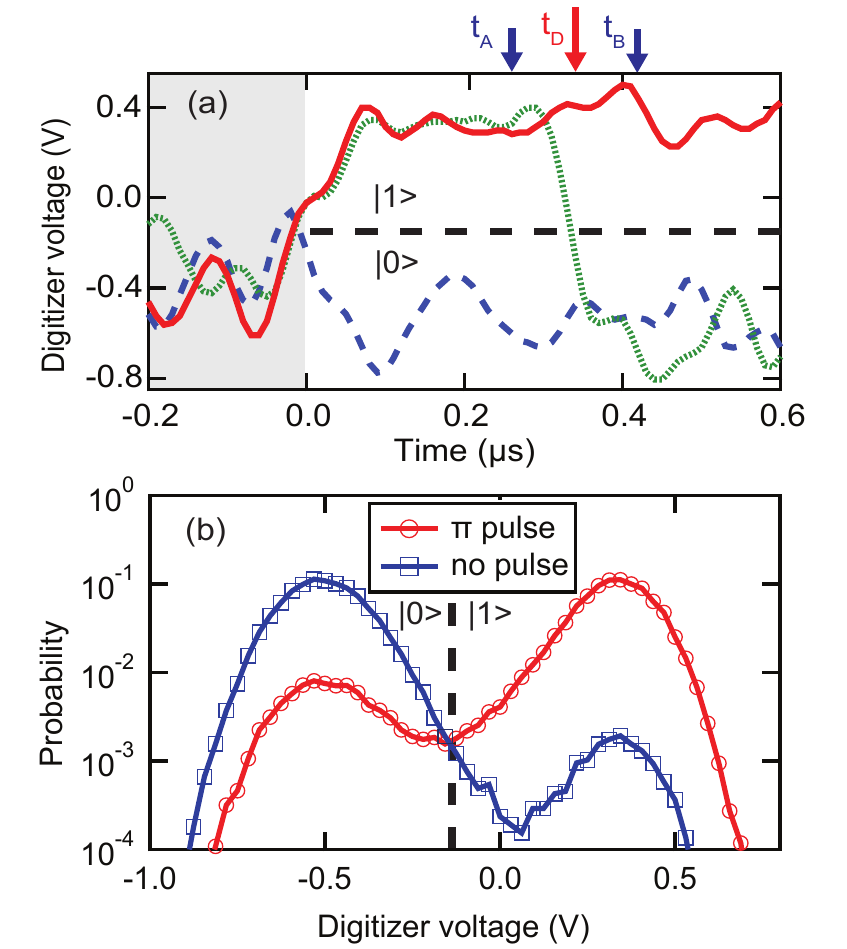}
\caption{(Color online) Qubit readout at $\bar{n}$ = 14.6 photons. (a) Three individual qubit readouts starting at $t = 0$: excited state (solid red line), ground state (long blue dashed line), abrupt quantum jump from the excited to ground state (short green dashed line). (b) Log-linear distributions of many ground and excited state preparations.}
\label{fig:jumps_and_hist}
\end{figure}

We first determine the measurement SNR, in particular, the degree to which we can separate the pointer state distributions. The readout strength is characterized by $\bar{n}$, the average photon population of the cavity. Figure \ref{fig:jumps_and_hist}(a) shows a portion of the readout signal as a function of time for three separate measurement traces at $\bar{n} =$ 14.6 photons. The dashed line indicates the discrimination threshold voltage for associating the cavity response with the ground or excited state of the qubit. The system SNR is sufficiently high to enable single-shot discrimination of the readout pointer states in a single 10-ns integration bin. Quantum jumps from the excited to ground state of the qubit are observed as an abrupt decrease in the signal. The characteristic decay time extracted from an exponential fit to the distribution of downward jump times agrees with the value of $T_1$ obtained from the ensemble averaged measurements.

In Fig. \ref{fig:jumps_and_hist}(b), we present distributions of many ground and excited state preparations calculated immediately after the cavity has come to equilibrium (90 ns into readout). The presence of the smaller, secondary peak in each distribution indicates partial contamination of the pure states and leads to fidelity loss. The distributions are well separated but there is a small number of errant counts, that is, counts at values of the homodyne voltage not centered about the bimodal peaks.  We attribute these counts to qubit transitions, induced by noise or $T_1$ relaxation, during the measurement.

\begin{figure}
\includegraphics[width=8.6cm]{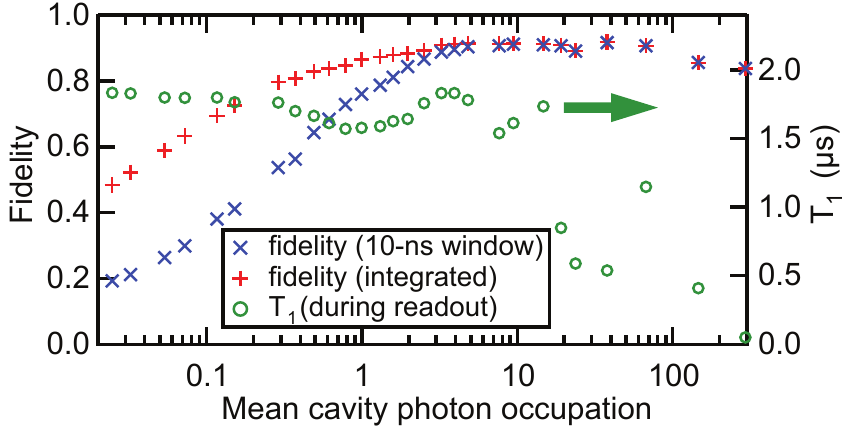}
\caption{(Color online) Raw fidelity and $T_1$ during readout as a function of mean cavity photon occupation.}
\label{fig:fidelity}
\end{figure}

We use these state distributions to calculate the readout fidelity $F=1-P_{0}-P_{1}$, where $P_0$ and $P_1$ are the fraction of error counts observed for a fixed discrimination threshold voltage. The uncorrected readout fidelity as a function of readout power is presented in Fig. \ref{fig:fidelity}. The ``10-ns'' fidelity series is calculated from the distributions of the homodyne voltage values from a single 10-ns digitization bin, as described above.  The ``integrated'' data series is calculated by integrating the entire readout signal weighted with an exponentially decaying filter \cite{exponential_filter}. The optimal filter time constant is empirically determined for each readout power to maximize the fidelity. At very low power, the 10-ns fidelity is severely suppressed because the system SNR is low. A substantial fraction of the lost fidelity can be recovered through integration. As expected, fidelity increases with increasing readout power and the two methods produce identical results above $\bar{n} \approx 10$, where the 10-ns readout distributions become well-separated.  The highest measured fidelity is $91.1 \pm 0.4 \%$ at $\bar{n} = 37.8$. In Fig. \ref{fig:fidelity}, we also plot the qubit relaxation time $T_1$ of the ensemble averaged signal during readout after an excited state preparation. Below $\bar{n} = 14.6$, $T_1$ during readout always exceeds 1.5 $\mu$s, but as the power is increased above $\bar{n} = 100$ both $T_1$ and the fidelity decrease as a result of increasing readout backaction.

\begin{figure}
\includegraphics[width=8.6cm]{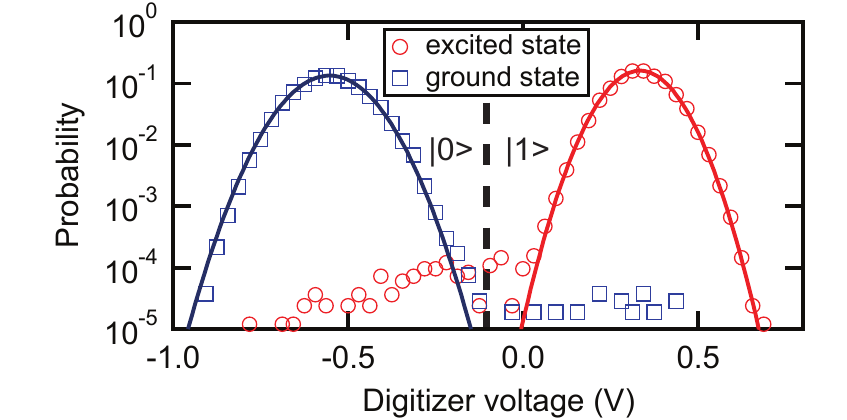}
\caption{(Color online) Pure readout pointer state distributions. A two-point correlation procedure is used to exclude qubit transitions and create pure ground and excited state distributions.}
\label{fig:filtered_hist}
\end{figure}

We now examine the measurement distributions in detail to investigate sources of fidelity loss. In particular, we assess the degree to which the pointer state distributions can be separated. In Fig. \ref{fig:jumps_and_hist}(b), we postulated that counts found between the two readout pointer state peaks were due to quantum jumps during measurement. To discard these events and create pure-state distributions, we implement a two-point correlation procedure. The system is probed at times $t_A$ and $t_B$, as shown in Fig. \ref{fig:jumps_and_hist}(a). We retain measurements only when both of these readings return the same value for the qubit state. Thus, the readout exhibiting the abrupt quantum jump is excluded. The 160-ns time difference between these points is several times longer than the response time set by the $\sim 10$~MHz system bandwidth, thus ensuring minimal autocorrelation between the signals at $t_A$ and $t_B$, and nearly random distributions at $t_D$. The resulting distributions at $t_D$ with these outliers removed are shown in Fig. \ref{fig:filtered_hist} for $\sim 10^5$ ground and excited state traces. Using the discrimination threshold shown in the figure, we observe only 108 false counts, thus allowing the pointer state distributions to be separated to one part in 1000. Consequently, the finite measurement SNR does not contribute to the observed fidelity loss. This procedure, however, does not correct for those rare events in which the system jumps twice. Unambiguously distinguishing these events is difficult due to the limited system bandwidth, which could be increased in future experiments.

\begin{figure}
\includegraphics[width=8.6cm]{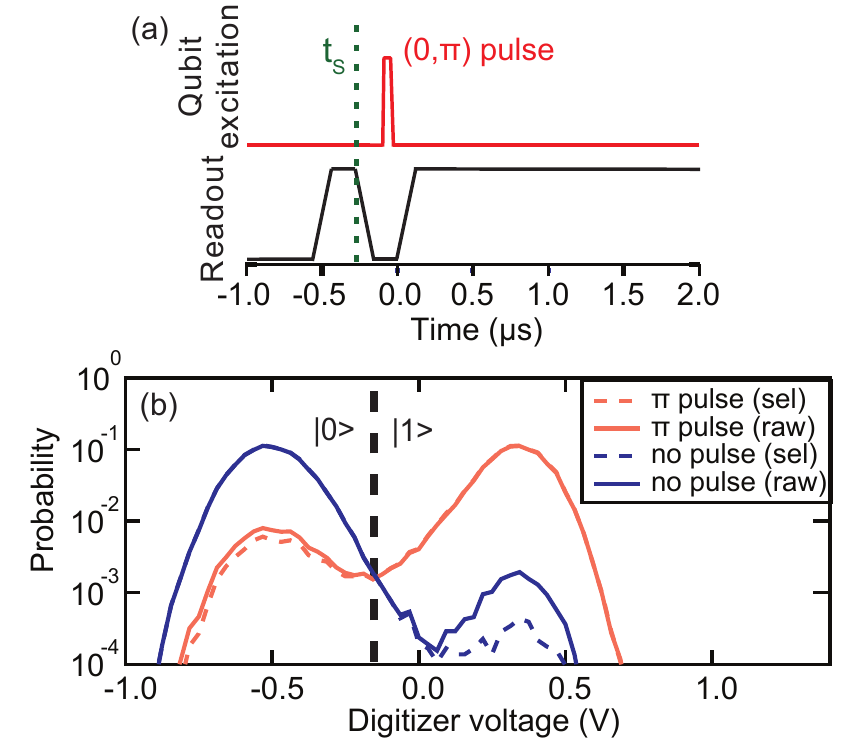}
\caption{(Color online) Heralded state preparation at $\bar{n}$ = 14.6. (a) Pulse sequence used for heralded state preparation. (b) Log-linear raw readout distributions for $\sim 10^5$ excited and ground state readout events (solid lines) compared with the distributions generated after heralded state preparation (dashed lines). The fidelity increases from 91.0\% to 93.9\%.}
\label{fig:selection}
\end{figure}

To investigate other sources of fidelity loss, we herald the ground state by inserting a fast measurement pulse prior to qubit manipulation, as shown in the pulse sequence of Fig. \ref{fig:selection}(a). With the qubit in thermal equilibrium with the environment, we energize the readout and extract the qubit state at $t_S$ (subscript $S$ denotes selection). If the qubit is determined to be in the ground state, no correction is applied; otherwise the subsequent readout is discarded from the total record. Events where the qubit is spontaneously found in the excited state can be attributed to either remnant thermal population or readout-induced excitation of the ground state \cite{dressed_dephasing_2009,slichter_DDarxiv} during the heralding pulse. With this procedure in place, we again prepare ground and excited state distributions, and observe a fidelity improvement of $2.9\pm0.2\%$. We compare the log-linear readout distributions both with (selected) and without (raw) heralded state preparation in Fig. \ref{fig:selection}(b). The reduced overlap of the distributions with heralded state preparation is demonstrated by the decreased size of the smaller of the bimodal peaks in both the excited and ground state distributions. This procedure corrects for the remnant thermal population of the excited state (88 mK effective temperature), which we attribute to stray infrared radiation and spurious electrical noise at the qubit frequency.

To separate out the contributions of the remaining sources of fidelity loss, we examine long time traces ($\gg T_1$) with the system prepared in the ground state. We record and analyze the statistics of individual transitions between the qubit states, extracting the average readout-induced excitation ($\Gamma_{\uparrow}$) and decay ($\Gamma_{\downarrow}$) rates. The calculated fidelity loss during measurement due to $\Gamma_{\uparrow}$ is $0.2\%$, while the contribution of $\Gamma_{\downarrow}$ is negligible. The remaining, unaccounted-for sources of fidelity loss are estimated to contribute 1.5\%. We speculate that this loss is explained by errors in excited state preparation, namely $\pi$-pulse imperfections and drifts of the Larmor frequency due to local magnetic flux variations. In Table \ref{tab:fidlosstab}, all of the sources of loss and the method by which they are calculated are listed. We note that the dominant loss mechanism is energy relaxation, and calculate its contribution by comparing the readout time and the measured value of $T_{1}$. Currently, there are flux qubits with demonstrated relaxation times on the order of 10 $\mu$s \cite{Bylander:2011,Steffen_CSFQ}; in our system, a similar $T_1$ would reduce the associated fidelity loss to 0.7\%. This level of coherence, coupled with heralded ground state preparation, high-precision $\pi$ pulses \cite{MIT_pulse_calibration}, and stable magnetic flux, should readily enable readout fidelities in excess of 98\% within this architecture.

\begin{table}[htbp]
\begin{ruledtabular}
\caption{A budget of fidelity loss at $\bar{n} = $ 14.6. The measured fidelity is $91.0 \pm 0.4\%$.}
\label{tab:fidlosstab}
\begin{tabular}{ c  c  c  }
Source of loss & Fidelity loss (\%) & Calculation method \\
\hline
$T_1$ decay & $4.4\pm0.3$ & measured $T_1$ \\
Thermal population & $2.9\pm0.2$ & heralding \\
$\Gamma_{\uparrow}$ & $0.2\pm0.1$ & individual jumps \\
$\Gamma_{\downarrow}$ & $0.0\pm 0.2$ & individual jumps \\
SNR & $<0.1$ & pointer state overlap \\
Remaining & 1.5 &  \\
\end{tabular}
\end{ruledtabular}
\end{table}

Finally, we propose a fast qubit initialization procedure based on heralding the ground state. If the measurement yields the excited state, this state could be rapidly reset to the ground state with a $\pi$ pulse using fast electronics. This approach would eliminate the need to change the detuning $\delta$ \cite{Yale_reset} to induce Purcell relaxation. The reset fidelity is ultimately limited by the raw measurement fidelity which includes the effect of $T_1$. We predict that for qubits with $T_1 >$  10 $\mu$s, a single iteration of this procedure would achieve reset with a fidelity close to 99\%, assuming state preparation errors could be eliminated.

In conclusion, we have demonstrated a fast, analog cQED readout to measure the state of a flux qubit with high single-shot fidelity. We use the QND nature of this readout at low powers to demonstrate heralded state preparation. This procedure allows us to obtain a detailed account of the sources of fidelity loss in the readout protocol. Moreover, errors due to imperfect ground state initialization can be readily identified and effectively eliminated through post-selection. Extending this further, these errors could be actively corrected in real time if a $\pi$ pulse were to be triggered based on the outcome of the heralding readout. The dominant remaining source of fidelity loss in our experiment is due to the short energy relaxation time of the qubit. For $T_1 >$ 10 $\mu$s with perfect state preparation, we predict a fidelity of greater than 98\% in the current readout architecture. We also note that this simple heralding procedure immediately allows for rapid reset of the qubit without the need to change bias parameters. Our readout architecture provides a new tool to study readout-induced backaction in a qubit with a high degree of anharmonicity, complementing previous work on transmon qubits \cite{dressed_dephasing_2009,slichter_DDarxiv}.

\begin{acknowledgments}
We thank E. M. Hoskinson and S. Weber for useful discussions and contributions to this project. D.H.S. acknowledges support from a Hertz Foundation Fellowship endowed by Big George Ventures. This work was funded in part by the U.S. Government and by BBN Technologies. Funding is also acknowledged from the ARO under grant W911NF-11-1-0029.

\textit{Note added.}\textemdash We note the complementary work of D. Rist\`{e} \textit{et al.} \cite{dicarlo_arxiv} that was recently brought to our attention.
\end{acknowledgments}

% References

% \bibliographystyle{jed_et_al}
% \bibliography{fluxqubitmsacitations}


\end{document}